\documentclass[twocolumn]{aastex631}
\newcommand\un[1]{{\,\rm #1}}
\newcommand\E[1]{\times10^{#1}}
\newcommand\rs[1]{_\mathrm{#1}}
\newcommand\g{$\gamma$}

\begin{document}

\title{Evidence of gradients of density and magnetic field in the remnant of Tycho's supernova}

\author[0000-0003-3487-0349]{Oleh Petruk}
\affiliation{INAF-Osservatorio Astronomico di Palermo, Piazza del Parlamento 1, 90134, Palermo, Italy}
\affiliation{Institute for Applied Problems in Mechanics and Mathematics, Naukova Street 3-b, 79060 Lviv, Ukraine}

\author[0000-0001-8464-0360]{Taras Kuzyo}
\affiliation{Institute for Applied Problems in Mechanics and Mathematics, Naukova Street 3-b, 79060 Lviv, Ukraine}

\author[0000-0002-3226-3118]{Mariana Patrii}
\affiliation{Faculty of Physics, Ivan Franko National University of Lviv, Kyryla and Methodia  8, 79005 Lviv, Ukraine}

\author[0000-0002-8400-3705]{Laura Chomiuk}
\affiliation{Center for Data Intensive and Time Domain Astronomy, Department of Physics and Astronomy, Michigan State University, East Lansing, MI 48824, USA}

\author[0000-0002-7918-904X]{Maria Arias}
\affiliation{Leiden Observatory, Leiden University, PO Box 9513, 2300 RA, Leiden, The Netherlands}

\author[0000-0003-0876-8391]{Marco Miceli}
\affiliation{Dipartimento di Fisica e Chimica E. Segr\`e, Universit\`a degli Studi di Palermo, Piazza del Parlamento 1, 90134, Palermo, Italy}
\affiliation{INAF-Osservatorio Astronomico di Palermo, Piazza del Parlamento 1, 90134, Palermo, Italy}

\author[0000-0003-2836-540X]{Salvatore Orlando}
\affiliation{INAF-Osservatorio Astronomico di Palermo, Piazza del Parlamento 1, 90134, Palermo, Italy}

\author[0000-0002-2321-5616]{Fabrizio Bocchino}
\affiliation{INAF-Osservatorio Astronomico di Palermo, Piazza del Parlamento 1, 90134, Palermo, Italy}

\begin{abstract}
By using surface brightness maps of Tycho's supernova remnant (SNR) in radio and X-rays, along with the properties of thermal and synchrotron emission, we have derived the post-shock density and magnetic field strength distributions over the projection of this remnant. Our analysis reveals a density gradient oriented towards the north-west, while the magnetic field strength gradient aligns with the Galactic plane, pointing eastward.
Additionally, utilizing this magnetic field map, we have derived the spatial distributions of the cut-off frequency and maximum energy of electrons in Tycho's SNR. 
We further comment on the implications of these findings for interpreting the gamma-ray emission from Tycho's SNR.
\end{abstract}


%--------------------------------------------------------------------
\section{Introduction} 
\label{tychograd:intro}

The remnant of Tycho's supernova (SN1572, G120.1+1.4) is observed at all wavelengths of the electromagnetic spectrum and 
marks the site of a 
type Ia supernova that appeared in 1572 \citep{2008Natur.456..617K}.
Cosmic rays are accelerated in Tycho's supernova remnant (SNR) to high energies \citep{2012A&A...538A..81M}. 
The availability of a significant amount of observational data over the past decades  makes this object one of the most interesting research targets.

Multiple radio and X-ray observations indicate that the expansion of the remnant into the interstellar medium (ISM) is anisotropic \citep{1997ApJ...491..816R, 2010ApJ...709.1387K, 2016ApJ...823L..32W, 2021ApJ...906L...3T}. The expansion parameter 
$(\Delta R/\Delta t)/(R/t)$ 
for the forward shock radius $R$ in Tycho's SNR varies between 0.2 and 0.8 as measured in radio or X-rays \citep{1997ApJ...491..816R,2010ApJ...709.1387K}. Thus, there should be irregularities in the ISM density that cause such variations in the proper motion of the shock. 

There have been multiple attempts to identify an interaction between the remnant and dense clouds in its vicinity by considering stronger shock deceleration \citep{1999AJ....117.1827R, 2004ApJ...605L.113L, 2016ApJ...826...34Z, 2017A&A...604A..13C}, or by measuring density variations around the shock front, which could be interpreted as the presence of  local ISM gradients \citep{2011ApJ...727...32V, 2013ApJ...770..129W}. There are a number of studies that have addressed the distribution of dense molecular material around the location of Tycho's SNR \citep{1999AJ....117.1827R,2011RAA....11..537X, 2004ApJ...605L.113L,2009ChA&A..33..393C, 2010A&A...521L..61I,2017A&A...604A..13C}.

Spatially resolved X-ray spectroscopy enables detailed ejecta velocity measurements from Doppler shifts in the source's emission lines \citep{2010ApJ...725..894H, 2017ApJ...840..112S, 2022ApJ...937..121M, 2023A&A...680A..80G,2024ApJ...962..159U}. It also reveals asymmetries in expansion of Tycho's SNR, and allows us to constrain the ejecta structure.

Little is known about the interstellar magnetic field (ISMF) structure in the remnant's surroundings.
Several authors impose limits on the magnetic field (MF) strength in the remnant based on observational data \citep[e.g.][]{2007ApJ...665..315C, 2020A&A...639A.124W,2021ApJ...917...55R}.
Radio and X-ray polarization studies  \citep{1997ApJ...491..816R,2023ApJ...945...52F} indicate a predominantly radial MF orientation with an amplification/compression factor of about 3.4 at the shock front.

Three-dimensional magneto-hydrodynamic (3D MHD) simulations of SNR evolution are necessary for a better understanding of the SNR structure, the physical processes related to the supernova explosion and its expansion into a non-uniform ISM, and for testing hypotheses about the cosmic ray acceleration and its high-energy emission.  
There are several HD or MHD models of Tycho's SNR that aim to explain its observational features as well as azimuthal asymmetries \citep{2013MNRAS.435.1659C,2017ApJ...842...28W, 2018MNRAS.474.2544F, 2019ApJ...877..136F,2020MNRAS.494.1531M,2024ApJ...961...32K}. 
The structure of the ISM density and the MF are crucial inputs for MHD models. Indeed, the ISM density is one of the key factors affecting the SNR shock wave dynamics. On the other hand, the MF is crucial for the non-thermal emission from SNRs, and for the cosmic ray acceleration. It could be important, too, during the interaction of the remnant with circumstellar medium or ISM inhomogeneities. It can also damp HD instabilities that can fragment the inhomogeneities after the shock passage. Furthermore, the MF is responsible for the (partial) suppression of thermal conduction \citep{2008ApJ...678..274O}.

The goal of this paper is to provide observation-based evidence for the large-scale gradients in both ISM density and ISMF strength that have direct implications for the evolution of Tycho's SNR. As a next step, we produce images of the distribution of the cut-off frequency and the maximum energy of electrons over the SNR.

%--------------------------------------------------------------------
\section{Observations and data reduction} 
\label{tychograd:obs}

We have used the 1.4~GHz map produced by \citet{2016ApJ...823L..32W} from Karl G. Jansky Very Large Array (VLA) observations of Tycho's SNR. The data were taken in 2013--2014 from a combination of A, B, C, and D configurations, observed under project code VLA/13A-426 (PI J.~W.\ Hewitt). Data from two spectral windows covering 1314--1442 GHz were used to image the SNR using uniform weighting. The resulting L-band map has a resolution of 1.91$''$, and a pixel size of 0.4$''$. 

The X-ray maps in the photon energy ranges $1.2-4.0$ keV, $4.1-6.0$ keV and $3.0-6.0$ keV are derived from \textit{Chandra} data taken in 2015 (obsid 15998). They were analyzed with the Chandra Interactive Analysis of Observations (CIAO) v.4.12 software. The data were reprocessed with \verb+chandra_repro+ and X-ray maps were produced with the  \verb+fluximage+ script. The derived X-ray maps have a resolution and pixel size of 0.492$''$.
The radio and X-ray images are taken at almost the same epoch, and therefore no correction is needed to account for the SNR expansion between the observations. 
The FITS files of the X-ray images are reprojected to the same pixel grid as the radio map to allow for a pixel-to-pixel analysis. This is done by using the flux-conserving function \verb+reproject_exact+ from the \verb+astropy+ library. 

For the spatial variation of the radio spectral index $\alpha$, we have used the $40''$ map derived by \citet{2019AJ....158..253A}. It was made by combining LOw Frequency ARay \cite[LOFAR;][]{2013A&A...556A...2V} data at 48.3~MHz, 67.0~MHz, and 144.6~MHz with VLA data at 327~MHz \citep{2000ApJ...529..453K} and 1382~MHz \citep{2016ApJ...823L..32W}. The spectral index map was produced by, for each pixel, fitting for a value of the flux density, spectral index, and line-of-sight absorption. The spectra index $\alpha$ (where flux density $S_{\nu} \propto \nu^{-\alpha}$) varies in value from $0.55$ to $0.75$ over the surface of Tycho's SNR. 
The pixel size in the the spectral index map is larger than in the radio and X-ray maps. Therefore, we prescribed the same radio spectral index for all pixels in our grid which correspond to a single one in the spectral index image. 

%--------------------------------------------------------------------
\section{Results}
\label{tycho3Da:sect-2D}

\subsection{Density and magnetic field strength}
\label{tycho3Da:sect-2D-1}

Our goal is to understand the large-scale distribution of the density and MF strength around Tycho's SNR. We have adopted our method for processing maps of SNRs in different photon energy ranges \citep{2009MNRAS.399..157P}. The basic idea is to derive images of the global distribution of density and MF strength over Tycho's SNR resulting from the evolution of the plasma inside (including both components: the swept-up ISM and the ejecta) after impact with the ambient medium. 

%%%%%----------------------------------------%%%%%
\begin{figure*}
  \centering 
  \includegraphics[width=0.495\textwidth]{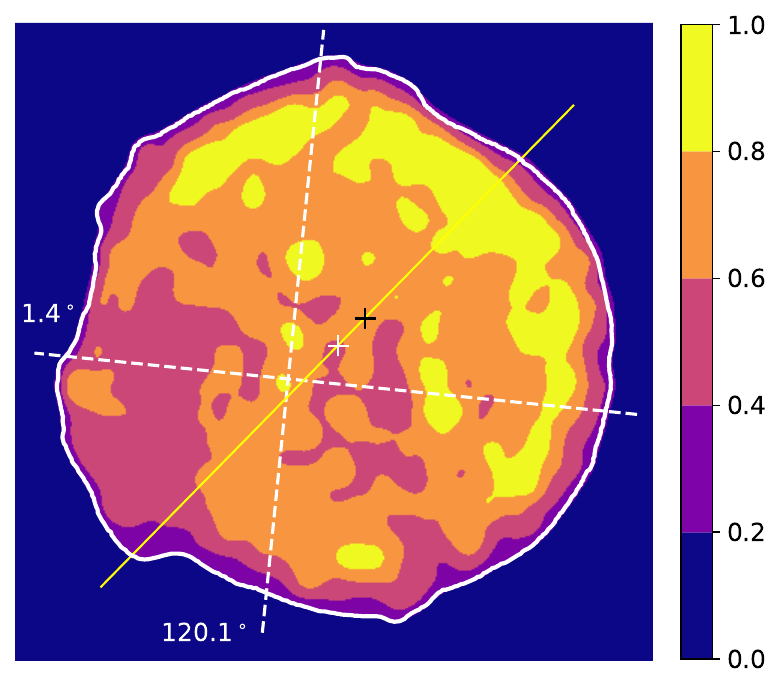}
  \includegraphics[width=0.495\textwidth]{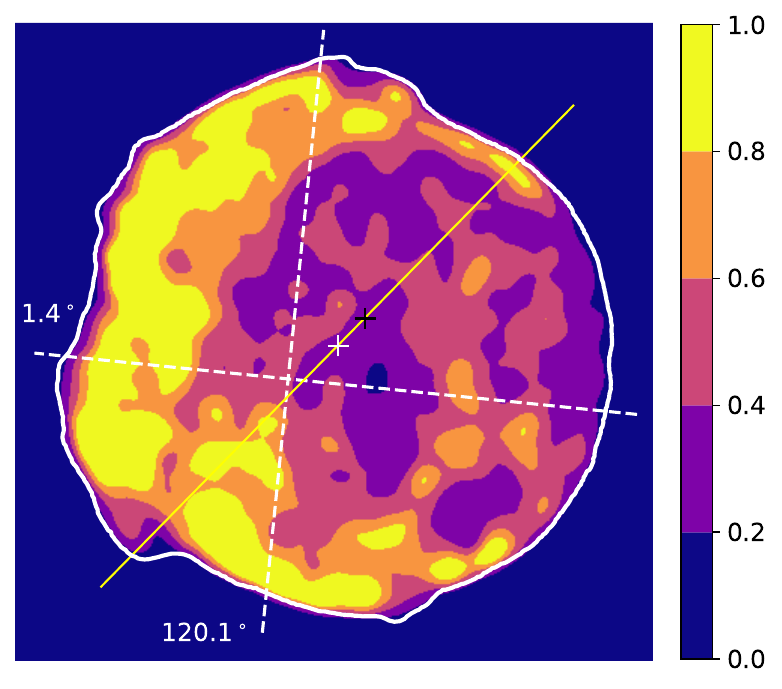}
  \caption{
   \textit{Left:} Density distribution derived from the thermal X-ray image of Tycho's SNR and equation (\ref{tycho3Da:eq_n}). \textit{Right:} Distribution of MF strength derived from the radio and thermal X-ray images of Tycho's SNR and equation (\ref{tycho3Da:eq_B}). The color scales are in arbitrary units. The outer contour of the SNR (white line) is produced from the radio image. 
   The white cross corresponds to the geometrical centre of the SNR, while the black one corresponds to the actual location of the explosion. The yellow line passes through the two centres. The white dashed lines mark Galactic coordinates $l=120.1\degr$, $b=+1.4\degr$. The Galactic plane is located below the remnant. The grid of equatorial coordinates is parallel to the edges of each figure;  north is on top, east on the left.
   Both images (with size  $1500\times1500$ pixels) are smoothed with a Gaussian ($\sigma \approx$ 20 pixels) to reduce small-scale fluctuations. 
  }
  \label{tycho3Da:fig_nBb}
\end{figure*}
%%%%%----------------------------------------%%%%%

The synchrotron radio emissivity is:
\begin{equation} 
 \epsilon\rs{r}\propto KB^{(s+1)/2},
 \label{tycho3Da:qradio}
\end{equation}
where $K\propto n$ is the normalization of the electron spectrum, $s=2\alpha+1$, $\alpha$ is the radio index, $n$ is the number density, and $B$ is the MF strength of the component in the plane of the sky. 
The surface brightness in each pixel of an image is the sum of the emissivity along the line of sight inside the SNR. 

We can write the radio brightness in each pixel as:
\begin{equation} 
 q\rs{r}\propto\eta\rs{r}nB^{(s+1)/2},
 \label{tycho3Da:qradio2}
\end{equation}
where $\eta\rs{r}$ is a geometrical factor, different for the different pixels in the image. It accounts for the distribution of the radio emission along the line of sight inside the SNR. 

The thermal X-ray emissivity is:
\begin{equation} 
 \epsilon\rs{x}\propto n^2 \Lambda(T,\tau),
 \label{tycho3Da:eq_qthermal}
\end{equation}
where $\Lambda(T,\tau)$ is a cooling function for optically thin plasma with temperature $T$ and ionization state as characterized by the parameter $\tau$. Tycho's SNR is a young object with a high shock speed $V$ (average $V\approx 3300\un{km/s}$ for distance $d=2.3\un{kpc}$), and high post-shock electron temperatures $T> 10^7\un{K}$ \citep{2013ApJ...770..129W}. 
%\op{\citet{2013ApJ...770..129W}: $T\rs{es}=1.6\E{7}\un{K}$, $T\rs{p}\simeq \un{K}$, $V\rs{ave}=3300\un{km/s}$ for $d=2.3\un{kpc}$.}
For the thermal X-ray map, we use the X-ray data in the photon energy range $1.2- 4\un{keV}$, where the function $\Lambda(T)$ is approximately constant for such temperatures and different ionization states (Appendix~\ref{Tycho3D:app1}). If this is the case, then the surface brightness in each pixel of the image is approximately:
\begin{equation} 
 q\rs{x}\propto \eta\rs{x}n^2,
 \label{tycho3Da:eq_qthermal2}
\end{equation}
where $\eta\rs{x}$ is a geometrical factor for the thermal X-ray emission.

Expression (\ref{tycho3Da:eq_qthermal2}) yields the density distribution  over the SNR projection:
\begin{equation} 
 n \propto (q\rs{x}/\eta\rs{x})^{1/2}.
 \label{tycho3Da:eq_n}
\end{equation}
The expression for the MF map over the SNR follows from equations (\ref{tycho3Da:qradio2}) and (\ref{tycho3Da:eq_qthermal2}): 
\begin{equation}
B \propto \left(\frac{q\rs{r}^2}{q\rs{x}}\frac{\eta\rs{x}}{\eta\rs{r}^2} \right)^{1/(s+1)}.
\label{tycho3Da:eq_B}
\end{equation}
By using $n$ from equation (\ref{tycho3Da:qradio2}) in equation (\ref{tycho3Da:eq_qthermal2}) we assume the same injection efficiency for electrons in the forward and reverse shocks. Differences in density between the shocked ISM and the ejecta material are reflected in the geometrical factors, which account for the internal structure of the remnant.

The geometrical factors $\eta\rs{r}$, $\eta\rs{x}$ vary from pixel to pixel. They  
may be estimated from geometrical considerations and the internal profiles of emissivity for the SNR. Namely, we use the Abel inversion:
\begin{equation}
    \eta\left(\overline{r}\right)=2\int_{\overline r}^{1}\overline f(r')\frac{r'dr'}{\sqrt{r'^2-\overline r^2}}
\end{equation}
where $\overline r=r/R$, $r$ is the distance from the explosion point, $R$ is the SNR radius, $\overline f(r)=f(r)/f\rs{s}$, $f\rs{s}\equiv f(R)$ is the post-shock value of the function $f(r)$, which represents the spatially variable part of the emissivity. We adopt the functions  $f\rs{r}=nB^{3/2}$ for the radio band and $f\rs{x}=n^2$ for the thermal X-rays. They are calculated using $n(r)$ and $B(r)$ from numerical simulations, namely, with the radial density and MF profiles for the remnant of an SN Ia explosion at age 450 yrs from our one-dimensional MHD simulations \citep{2021MNRAS.505..755P}. It appears that  $\eta\rs{r}\approx\eta\rs{x}$ for $\overline r=0.75-1$ (numerically, they are in the range $0.20-0.36$), and differ from one another by about $30\%$ for smaller radii $\overline r<0.75$ (with values between $0.12$ and $0.20$). We apply these radial functions $\eta\rs{r}(\overline r)$ and $\eta\rs{x}(\overline r)$ for different azimuths by scaling them with their respective $R$ for each direction from the explosion center. We take the coordinates of the SN center from \citet{2015ApJ...809..183X}. Note that $\eta$ is determined by using a normalization to the post-shock values ($\bar r=r/R$, $\bar f=f/f\rs{s}$). Therefore, equations (\ref{tycho3Da:eq_n}) and (\ref{tycho3Da:eq_B}) for the density and MF strength give not the average values along the line of sight, but rather the post-shock values of $n$ and $B$.

The derived images for the post-shock number density and the post-shock MF strength in Tycho's SNR are shown in Fig.~\ref{tycho3Da:fig_nBb}.
There are also two centres shown on these plots: the geometrical center \citep[which corresponds to the circular fit of the outer edge of the SNR;][]{2005ApJ...634..376W,2010ApJ...709.1387K}, and the actual center \citep[estimated from the backward extrapolation of the proper motion vectors for different filaments;][]{2015ApJ...809..183X}. 

%%%%%----------------------------------------%%%%%
\begin{figure}
  \raggedleft
  \includegraphics[width=0.99\columnwidth]{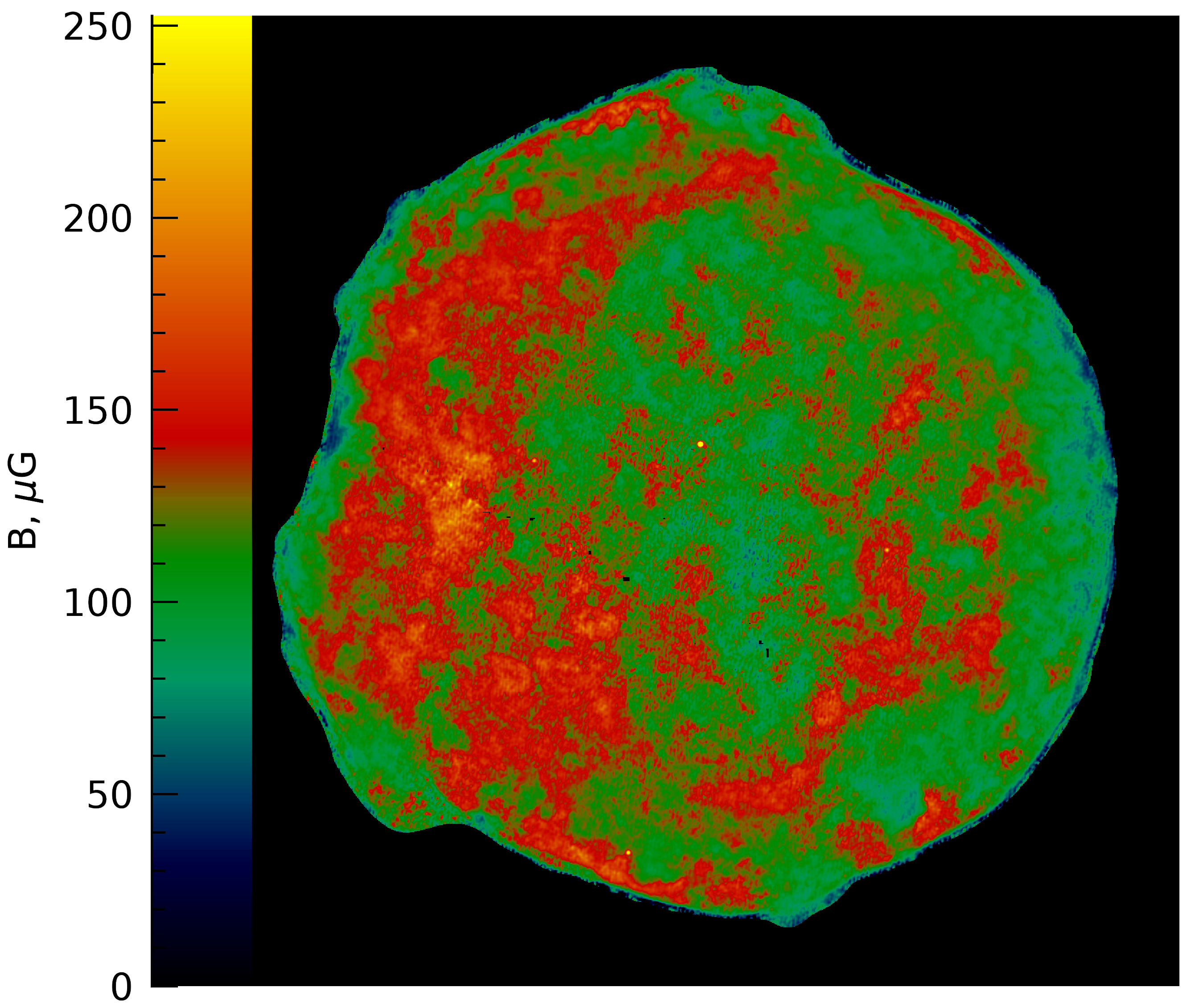}\\
  \includegraphics[width=\columnwidth]{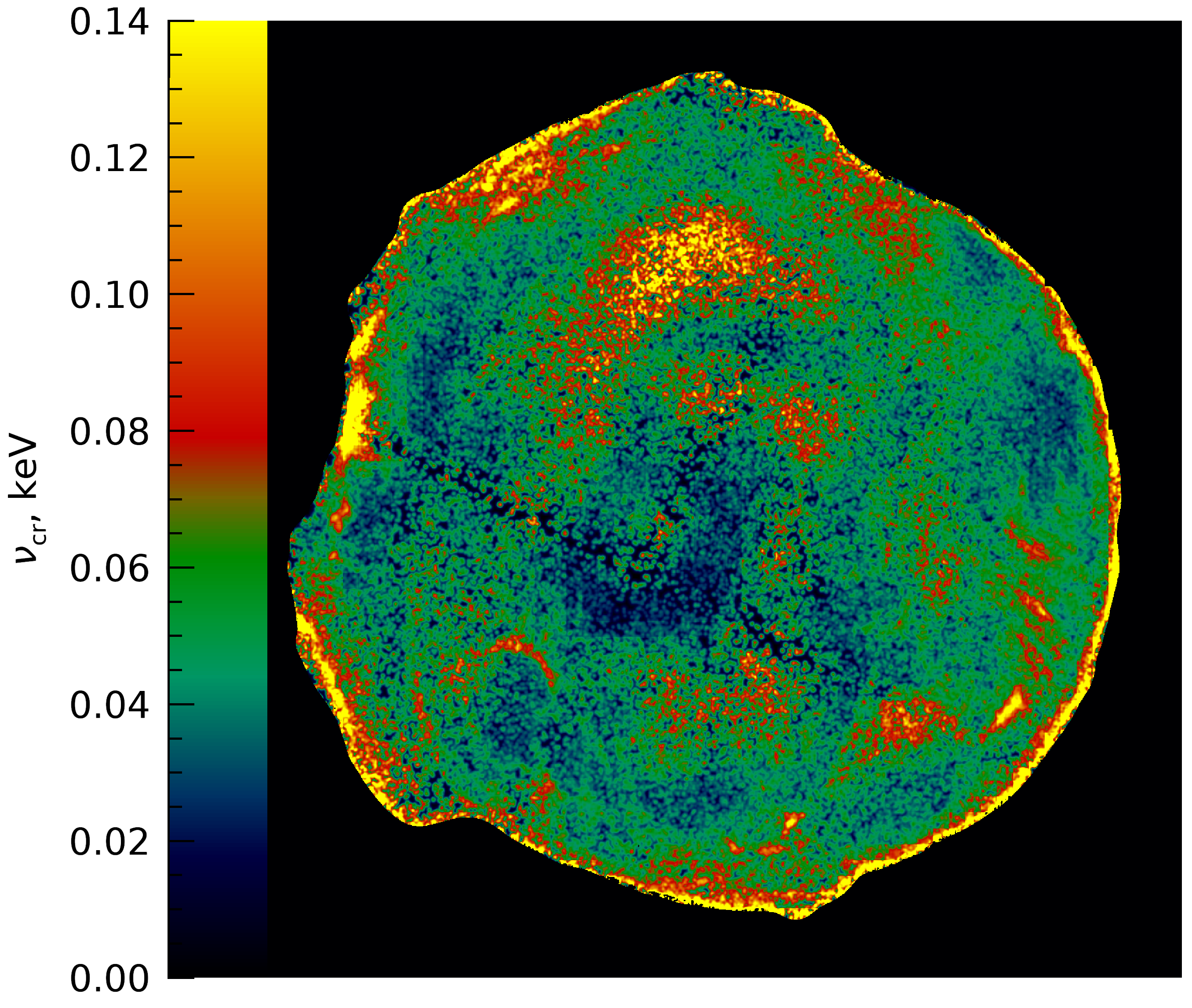}\\
  \includegraphics[width=0.95\columnwidth]{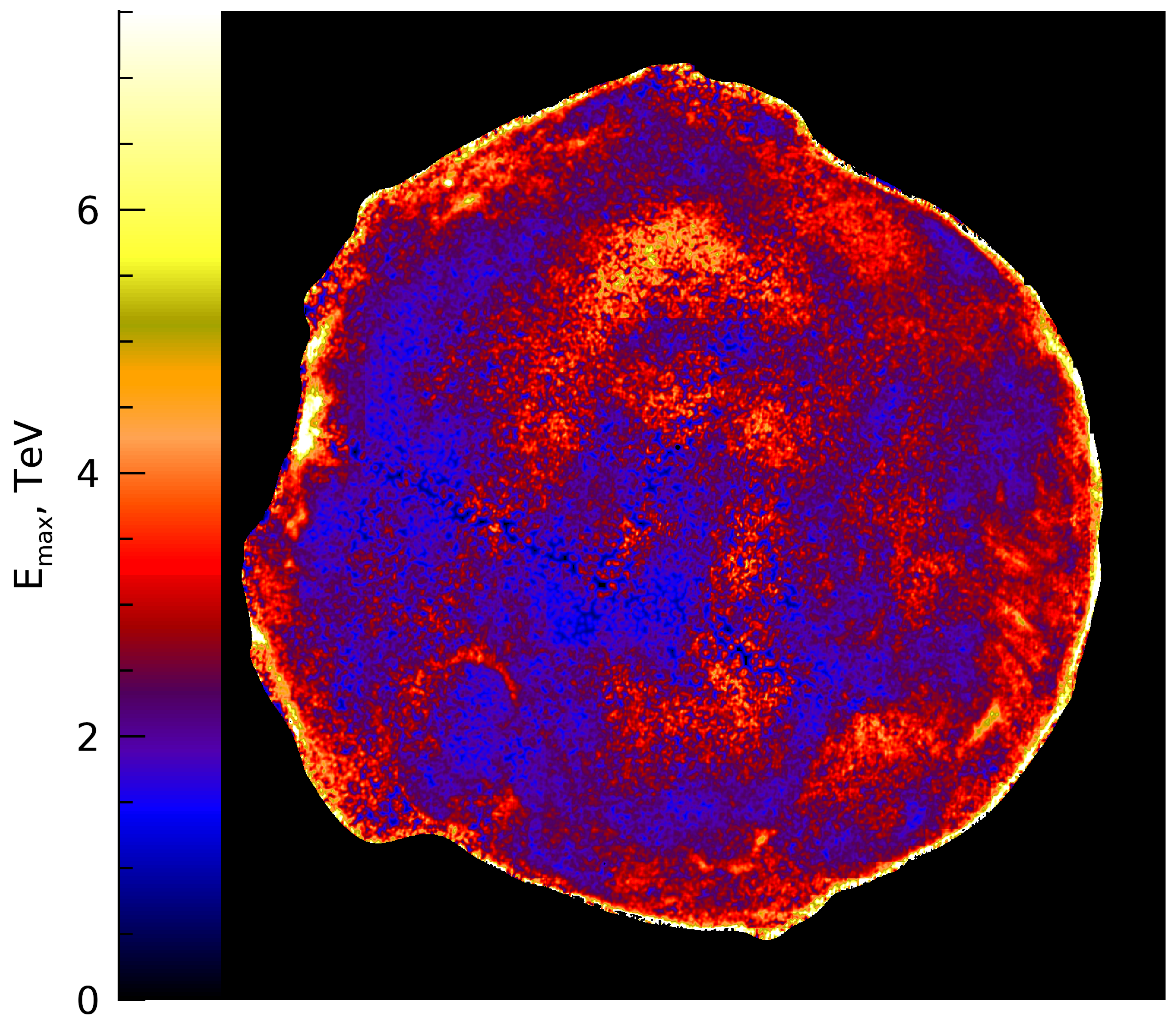}
  \caption{
   {\it Top:} MF strength $B$. The distribution of values of $B$ is close to a Gaussian with $120\pm23\un{\mu G}$.
   {\it Centre:} Cut-off photon energy $h\nu\rs{cr}$ derived from observational data with Eq.~(\ref{tycho3Da:eq_nubr}). 
   {\it Bottom:} Maximum energy of electrons $E\rs{max}$ as given by Eq. (\ref{tycho3Da:eq_Emax}). 
  }
  \label{tycho3Da:fig_nubr}
\end{figure}
%%%%%----------------------------------------%%%%%

\subsection{Cut-off frequency and maximum energy of the electrons}

Once we have the map of the MF, we may apply a procedure similar to one used in the \verb+srcut+ spectral fitting model in \verb+XSPEC+ and consider the particle acceleration. We use the fluxes at the radio $\nu\rs{r}$ and the X-ray $\nu\rs{x}$ frequencies as well as the radio spectral index to reconstruct the broadband synchrotron emission of the radiating electrons.

Tycho's SNR is a young object. The maximum energy of the electrons $E\rs{max}$ is likely limited by its age rather than radiative losses. A good proxy for the particle distribution in energy, in this case, is a power law with an exponential cut-off. 
The accurate approximation to the synchrotron spectrum corresponding to such a model is given by equation (8) in \citet{2009MNRAS.399..157P}:
\begin{equation}
 \epsilon\rs{xs}=\epsilon\rs{r}\left(\frac{\nu\rs{x}}{\nu\rs{r}}\right)^{-(s-1)/2}
 \exp\left[-\beta\rs{x}\left(\frac{\nu\rs{x}}{\nu\rs{cr}}\right)^{0.364}\right]
\label{tycho3Da:eq_apprsynch},
\end{equation}
where $\nu\rs{cr}=c_1BE\rs{max}^2$ is the cut-off frequency (i.e. the critical frequency for the electrons with the maximum energy), $\beta\rs{x}=1.46+0.15\cdot(2-s)$, $c_1 = 6.26\E{18}\un{cgs}$, and $\epsilon\rs{xs}$ refers to the X-ray synchrotron emission. 

We have to account for the geometrical factors when applying this relation to the maps: $q\rs{r}=\eta\rs{r}\epsilon\rs{r}$, $q\rs{xs}=\eta\rs{xs}\epsilon\rs{xs}$. 
It is known that the length scale for the radial distribution of the X-ray emitting electrons is shorter than for the radio emitting particles. The synchrotron X-ray emission is from electrons with energies around $E\rs{max}$. Therefore, their normalized density $\bar n=n/n\rs{s}$ downstream of the shock is about $e$ times smaller compared to the normalized density of electrons with a power-law spectrum.
Thus, the geometrical factor for the synchrotron X-rays is $\eta\rs{xs}\simeq e^{-1}\eta\rs{r}$. 
By solving equation (\ref{tycho3Da:eq_apprsynch}) for the cut-off frequency, we have:
\begin{equation}
 \nu\rs{cr} \simeq \nu\rs{x}\left(\frac{1}{\beta\rs{x}}\ln \left[ 
 \frac{\eta\rs{xs}q\rs{r}}{\eta\rs{r}q\rs{xs}}
 \left(\frac{\nu\rs{r}}{\nu\rs{x}} \right)^{(s-1)/2} \right]
 \right)^{-2.75}.
 \label{tycho3Da:eq_nubr}
\end{equation}
We use $\nu\rs{r}=1.4\E{9}\un{Hz}$, $\nu\rs{x}=1\E{18}\un{Hz}$ (corresponding to $4.1\un{keV}$) and, for $q\rs{xs}$, the hard X-ray image of Tycho's SNR in the $4.1-6$ keV range.

The definition of the critical frequency gives the expression to produce the map of the maximum energy of electrons from the maps of $B$ and $\nu\rs{cr}$: 
\begin{equation}
 E\rs{max}=\left(\frac{\nu\rs{cr}}{c_1B}\right)^{1/2}.
\label{tycho3Da:eq_Emax}
\end{equation}

The image of MF $\bar{B}$ on Fig.~\ref{tycho3Da:fig_nBb} is in normalized units. In order to convert it to physical units $B$, we calculate the mean value of the MF strength in normalized units $\langle \bar{B}\rangle$ and assume that this value corresponds to the mean field in the remnant $\langle B\rangle= 120\un{\mu G}$.
Fig.~\ref{tycho3Da:fig_nubr}, \textit{top}, shows the map of the MF where each pixel has strength $B=\bar{B}\cdot \langle B\rangle/\langle \bar{B}\rangle$. 
We see from this figure that the mean value which we adopted provides $B\simeq 200\un{\mu G}$ in the regions along the SNR rim; this value coincides with that estimated by other authors \citep[][and references therein]{2021ApJ...917...55R}. Thus, we substitute equation (\ref{tycho3Da:eq_Emax}) with $B$ calculated in this way. 

The images in Fig.~\ref{tycho3Da:fig_nubr} show the distribution of the cut-off frequency and the maximum energy of the electrons over Tycho's SNR obtained with this approach. 

Although synchrotron emission dominates in the regions around the edge of the SNR \citep{2015ApJ...814..132L}, thermal emission also contributes to the total flux in the hard X-rays. In order to be certain in our results, we calculate $\nu\rs{cr}$ by also considering the synchrotron-to-thermal X-ray fraction $\kappa$. By utilizing \textit{Chandra} and \textit{IXPE} observations, \citet{2023ApJ...945...52F} derived the synchrotron fraction map for the 3--6 keV photons over Tycho's SNR. In this photon energy range, the value of $\kappa$ varies from 0.16 to 0.90 across the SNR (their figure 7). We have used this map by setting $q\rs{xs}=\kappa q\rs{xh}$ where $q\rs{xh}$ corresponds to the observed X-ray map in $3-6$ keV. 
%As one can see from Fig.~\ref{tycho3Da:fig_nubr_frac} 
The images of $\nu\rs{cr}$ and $E\rs{max}$  derived in this way are the same as on Fig.~\ref{tycho3Da:fig_nubr} with a bit lower resolution due to the structure of the map for $\kappa$.

%%%%%----------------------------------------%%%%%
\begin{figure}
  \raggedleft
  \includegraphics[width=\columnwidth]{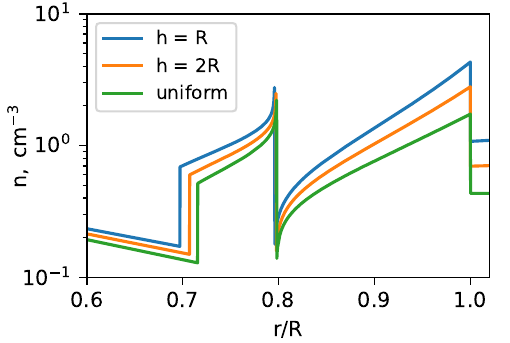}
  \caption{Density profiles corresponding to the remnants of an explosion in an ISM with density distribution (\ref{tycho3Da:eq_expdens}) with $h=R$ and $h=2R$, compared to the case of a uniform ISM density $n\rs{o}=\mathrm{const}$. $R$ is the SNR radius in the constant $n\rs{o}$. The profiles correspond to an age of 450 yrs and are the result of 1-D numerical simulations for an initial ejecta structure corresponding to SNIa \cite[see details in][]{2021MNRAS.505..755P}.
  }
  \label{tycho3Da:fig_dens_profiles}
\end{figure}
%%%%%----------------------------------------%%%%%

%-----------------------------------------------------------------
\section{Discussion and Conclusions}

\subsection{Gradients of density and magnetic field strength}

We have already noted at the end of Sect.~\ref{tycho3Da:sect-2D-1} that our density and MF maps were derived in such a way to give the post-shock values of $n$ and $B$. Although the ejecta material considerably contributes to the thermal X-ray emission in Tycho's SNR, the left map on Fig.~\ref{tycho3Da:fig_nBb} shows the large-scale distribution of the ambient density. 
The density of the shocked ejecta reflects the large-scale structure of the ISM because the flow structure adapts to the ambient conditions. Really, Fig.~\ref{tycho3Da:fig_dens_profiles} shows the profiles of the density behind a shock evolving in an ISM with density 
\begin{equation}
 n\rs{o}(r)=n\rs{o}(0)\exp(r/h),
 \label{tycho3Da:eq_expdens}
\end{equation} 
where $r$ is the distance from the explosion center and $h$ the scaling factor for the non-uniformity. Indeed, the ejecta density (between 0.7 and 0.8 $r/R$) is higher in the case of a stronger gradient. Another piece of information from this plot is that the different profiles are quite similar and may be scaled. This provides a justification for our procedure to use the geometrical factors $\eta$.

We have previously studied how the non-uniformity of interstellar density and interstellar MF affects the thermal X-ray \citep{1999A&A...344..295H}, and nonthermal radio \citep{2007A&A...470..927O}, X-ray and gamma-ray \citep{2011A&A...526A.129O} morphologies of SNRs. Comparing Fig.~\ref{tycho3Da:fig_nBb} with the general properties revealed in these references, we may come to the following conclusions about gradients. 

It is apparent from Fig.~\ref{tycho3Da:fig_nBb}, \textit{left}, that the density is higher in the north-west region. The yellow line is plotted to pass through the two centres (white and black crosses).  
It appears to mark the symmetry axis for the density distribution in Tycho SNR. In fact, this property provides strong evidence for the presence of a large-scale density gradient in the ISM around Tycho's SNR, which is directed towards the north-west, along the yellow line. Theoretical studies demonstrate \citep{1999A&A...344..295H} that the actual centre of the SNR should be shifted with respect to the geometrical one in the direction of the gradient of the ambient density. Thus, our figure is an independent confirmation of the explosion center found by \citet{2015ApJ...809..183X}.

There is `direct' evidence for a pre-shock density variation around Tycho SNR.
One piece of evidence comes from proper motion analysis of the forward shock in Tycho's SNR on the basis of radio observations \citep{1997ApJ...491..816R}. Figure 7 in this reference shows the azimuthal variation of the expansion parameter $m=(\Delta R/\Delta t)/(R/t)$. As from this study, $m$ marks prominent shock deceleration -- due to the interaction with increased density -- in the north-western direction. This is the same as the direction of the density gradient we found. 

In the same study, the parameter $m$ also demonstrates a strong shock deceleration towards the east, along the azimuth $\approx 70^\mathrm{o}$ (counterclockwise from the north). A considerable decrease of the expansion velocity at this azimuth is confirmed by X-ray data as well \citep{2016ApJ...823L..32W}. Analysis of infrared observations reveals a prominent increase of the pre-shock density in both the north-western and eastern directions \citep{2013ApJ...770..129W}.

In contrast, our map on Fig.~\ref{tycho3Da:fig_nBb} does not reflect a density enhancement towards the east. Our method does not `measure' the density around the shock. It instead relies on the internal distributions. 
Therefore, the difference between the two azimuths should be due to different spatial scales of the corresponding over-densities and time-scales of interactions. The one towards the north-west is large-scale (larger than the SNR radius). Therefore, the SNR evolved throughout its entire life in an ISM with this gradient, and the internal structure of the SNR has has enough time to adapt to such an ambient density variation. In contrast, the density enhancement on the east should be localized, and the interaction with the shock is small-scale. According to our interpretation, the shock has encountered it rather recently, so the downstream structure does not reflect this interaction yet, unlike the shock itself.

Recent deceleration of the shock at azimuth $\approx 70^\mathrm{o}$ is confirmed by the analysis of the proper motion of the forward shock from multi-epoch X-ray observations between years 2003 and 2015 \citep[][their region 4]{2021ApJ...906L...3T}. The authors found as well a prominent gradual decrease of the shock speed for the south and south-western edge of Tycho's SNR (their regions 6--10, azimuths $150-270$ degrees). There is a sign of such a behaviour (namely, smaller $m$) around the azimuth $180^\mathrm{o}$ in figure 7 in \citet{1997ApJ...491..816R}. 
\citet{2021ApJ...906L...3T} interpreted this as evidence of a recent interaction with a denser cavity wall. However, there is no excessive post-shock density found in the infrared 2004 data in the south and south-west \citep{2013ApJ...770..129W} that could result in such a rapid deceleration. Thus, in the scenario proposed by \citet{2021ApJ...906L...3T}, the south-west portion of the shock should start interaction with a higher-density feature in more recent times than the eastern portion around the azimuth $70^\mathrm{o}$ because the over-density at this azimuth is visible in the infrared data.

Fig.~\ref{tycho3Da:fig_nBb} \textit{right} demonstrates that the MF map in Tycho's SNR has an arch-like feature. In case of a uniform ambient MF, the MF configuration in the SNR is expected to be barrel-like due to a higher compression of the tangential MF component compared to the radial one. The MF projection would then have two symmetric limbs. Such an arch-like configuration in Tycho's SNR could appear if e.g. the 
two limbs are slanted toward the east and converging there. 
Such a configuration \citep{2007A&A...470..927O,2011A&A...526A.129O} may naturally happen if the gradient of the ambient MF is directed towards the east and it is just swept-up and compressed by the shock. Instead, if such a young SNR still has significant  amplification of the MF around the forward shock, then one would expect that MF upstream is approximately either:
\begin{equation}
 \delta B^2\propto n\rs{o}V^3,
\end{equation}
for non-resonant amplification, or:
\begin{equation}
 \delta B^2\propto B\rs{o}n\rs{o}^{1/2}V,
\end{equation}
for the resonant case. The compression of such a disordered MF by the shock is independent of its obliquity. Thus, in the first case, we  expect the MF map to look similar to the density map, which is not the case for ours, or to be dominated by the regions with the highest shock speed $V$, which are neither those on the north-west or east. The resonant amplification case could in principle result in a MF map as on Fig.~\ref{tycho3Da:fig_nubr} if the ordered ambient $B$ dominates the azimuthal variation of density and shock speed.

The symmetry axis of the MF distribution seems parallel to the white dashed line running from the west to the east (which indicates Galactic latitude $b=+1.4\degr$). Therefore, the ISMF gradient around Tycho's SNR is parallel to the Galactic plane. Interestingly, such orientation differs from the orientation of the MF gradient in the remnant SN1006, where it is directed towards the Galactic plane  \citep{2011A&A...531A.129B}.

We have no observational clues regarding the orientation of the ambient MF. The radio and X-ray polarisation maps reveal a mostly radial MF inside the SNR projection \citep{1997ApJ...491..816R,2023ApJ...945...52F}. This is a young SNR, and the generation of a turbulent MF should be efficient. Such a MF component can change considerably the MF direction compared to what we may expect in the case of a simple compression of the interstellar field.  

To summarise, it looks like the ambient environment around Tycho's SNR has the large-scale gradients: of the density in the north-west direction, and of the MF, which is oriented toward the east and is almost parallel to the Galactic plane.

\subsection{Cut-off frequency and maximum energy of the electrons}

The electrons emitting X-rays suffer from radiative losses. They lose energy rather quickly and the synchrotron X-ray rims of SNRs are therefore thin. We see this clearly in the distributions of the cut-off frequency and the maximum energy on Fig.~\ref{tycho3Da:fig_nubr}. The highest values are around the SNR edge, with a trend to somewhat larger cut-off frequencies and maximum energies in the western half of the remnant's rim. 

\begin{figure}
  \raggedleft
  \includegraphics[trim=0 14 32 30,clip,width=0.99\columnwidth]{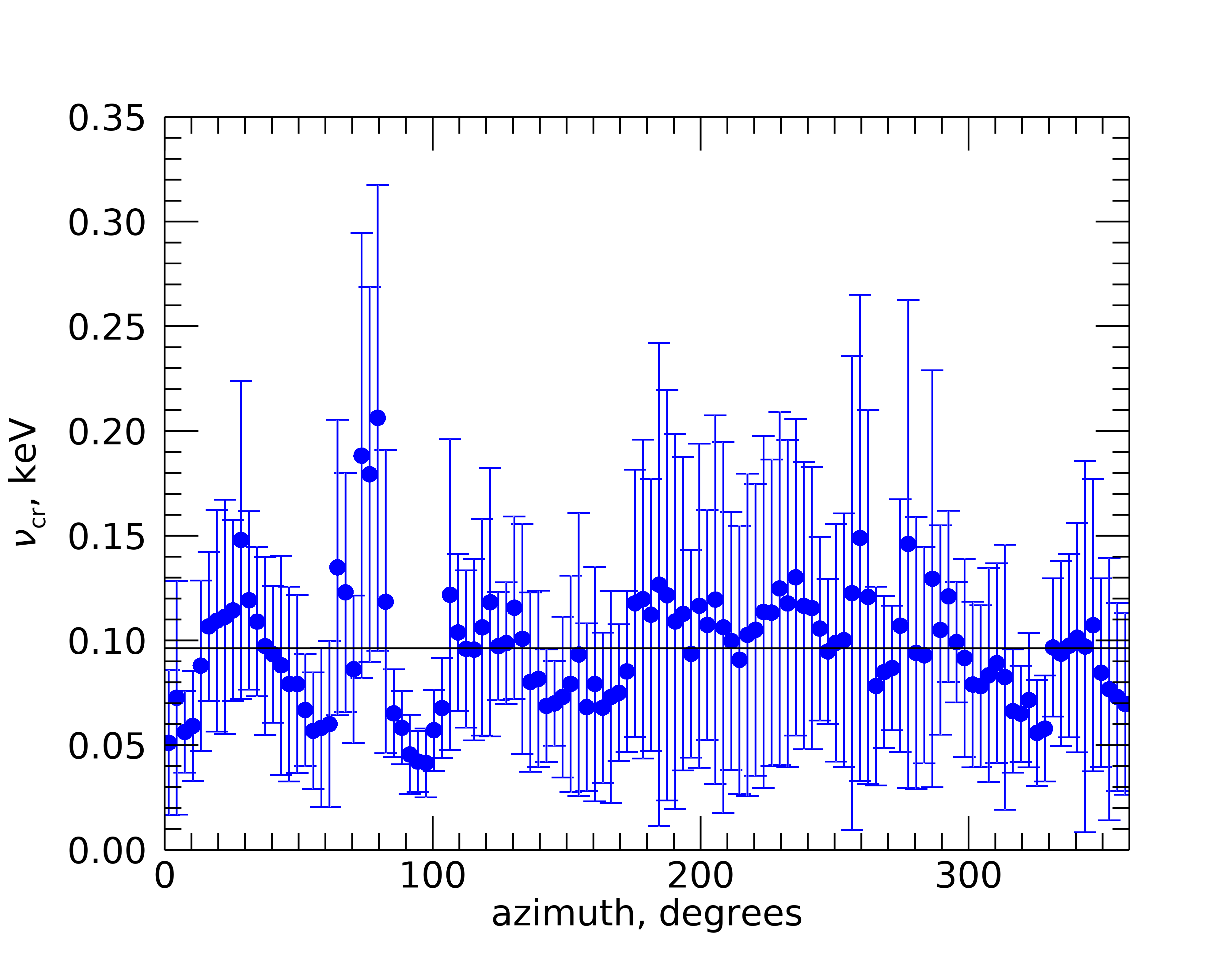}
  \caption{Mean and standard deviation of 
   $\nu\rs{cr}$ around the SNR edge. The azimuth is measured counterclockwise from the north. Each point on the plot is calculated with the data from a region limited by a $3$-degrees sector and $5\%$ extension inward along a radius from the edge. The black horizontal line shows the mean value for the whole rim (within $5\%$ of the radius).
  }
  \label{tycho3Da:fig_nucr_az}
\end{figure}

The azimuth variation of the cut-off frequency around the SNR edge is plotted on Fig.~\ref{tycho3Da:fig_nucr_az} for angular bins of 3 degrees. One of the most prominent enhancements is in the region of the recent shock-cloud interaction around azimuth $70^\mathrm{o}$. The local values reach up to $\nu\rs{cr}\simeq 0.42\un{keV}$ and $E\rs{max}\simeq 11\un{TeV}$ at this location. The variation of the cut-off frequency in other regions is typically within $50\%$ of the mean value for the whole rim, with evidence for values systematically above the mean in the western part of the shell (azimuth from $180$ to $290$ degrees, Fig.~\ref{tycho3Da:fig_nucr_az}). 

\citet{2015ApJ...814..132L} measured the cut-off frequency by fitting the NuSTAR spectrum in several regions over the remnant. Their figure 10 shows features similar to ours: in the interior, $\nu\rs{cr}$ is generally smaller by a factor of $\approx 2.5$ compared to the rim; it is higher over the western rim and locally in the east. They estimated, by assuming a uniform MF, that $E\rs{max}$ varies spatially in the range of $5-12\un{TeV}$. 
\citet{2006A&A...453..387P} estimated a MF strength of $B\approx 200\un{\mu G}$ from rim thickness as well as a maximum energy of electrons $E\rs{max}\simeq 5-7\un{TeV}$. 
These numbers agree with our calculations (Fig.~\ref{tycho3Da:fig_nubr}, \textit{bottom}).

On both our maps (for $\nu\rs{cr}$ and $E\rs{max}$, Fig.~\ref{tycho3Da:fig_nubr}), we see higher values also around well-known non-thermal features such as the stripes at the west ($E\rs{max}\simeq 4.0\un{TeV}$) and the arch in the south-east region ($E\rs{max}\simeq 3.7\un{TeV}$) that prove efficient energization of particles there.
The overall distribution of the maximum energy on our image is characterized by a mean $E\rs{max}=2.6\un{TeV}$ and a standard deviation $0.9\un{TeV}$. It is $E\rs{max}=4.0\pm1.7\un{TeV}$ for the rim.

%%%%%----------------------------------------%%%%%
%%%%%----------------------------------------%%%%%

\subsection{On the gamma-ray emission} 
\label{tychograd:gammas}

Tycho's SNR was detected in GeV \citep{2012ApJ...744L...2G} and TeV \citep{2011ApJ...730L..20A} \g-rays. Hadronic \g-rays arise from regions with a high density of target protons. There are two preferable locations in the case of Tycho's SNR, where the shock is interacting with denser ISM \citep{2013ApJ...770..129W}: in the north-west (in the direction of the large-scale gradient) and in the east (around azimuth $70^\mathrm{o}$). 

Neither VERITAS nor Fermi can resolve the object to distinguish between the two locations. Indeed, the centroid position for the TeV \g-ray emission is in the middle of the line between the two sites \citep[][their figure 1]{2011ApJ...730L..20A}. In the GeV \g-rays, the centroid is around the center of SNR \citep{2012ApJ...744L...2G,2017ApJ...836...23A}, i.e. is also between the two sites.

Our results could shed some light on differences in particle acceleration and \g-ray emission at these two locations. 
The shock speed $V$ in the north-west is around $3400\un{km/s}$ and near $2000\un{km/s}$ in the east \citep{2013ApJ...770..129W}. 
One would expect to have more efficient acceleration of cosmic rays in the north-west because of the higher shock speed. However, it follows from our results that $V$ measured in the east got smaller recently and sharply, in contrast to the north-west where the shock was decelerating continuously from early times due to increasing ambient density. Before the encounter with the dense material in the east, the shock speed could be as high as the highest $V$ in Tycho (which is around $4000\un{km/s}$; the same reference). The fact that $E\rs{max}$ for the electrons is higher in the east (being achieved before the sharp decrease of the shock speed due to a recent interaction with an over-density) than in the north-west (where the shock speed was lower for a long time) confirms such a scenario (Fig.~\ref{tycho3Da:fig_nubr} \textit{bottom}). Thus, we would favour the eastern region as the major site contributing to the \g-ray flux: cosmic rays were accelerated efficiently at the fast shock there and then `suddenly' hit the cloud with a high density of target protons.
This supports the scenario of particle acceleration in a shock interacting with dense medium described by \citet{2016A&A...589A...7M}.

%%%%%%%%%%%%%%%%%%%%%%%%%%%%%%%%%%%%%%%%%%%%%%%%%%%%%%%%
\begin{acknowledgments}
We are thankful to Adam Foster for valuable suggestions related to usage of the AtomDB data in numerical calculations as well as to Riccardo Ferrazzoli who provided us with the synchrotron fraction map for Tycho SNR. 
This paper employs a list of Chandra datasets, obtained by the Chandra X-ray Observatory, contained in~\dataset[DOI: 10.25574/cdc.259]{https://doi.org/10.25574/cdc.259}. 
O.P. acknowledges the OAPa grant number D.D.75/2022 funded by Direzione Scientifica of Istituto Nazionale di Astrofisica, Italy. 
L.C. is grateful for support from National Science Foundation grants AST-2107070 and AST-2205628.
This project has received funding through the MSCA4Ukraine project, which is funded by the European Union. Views and opinions expressed are however those of the authors only and do not necessarily reflect those of the European Union. Neither the European Union nor the MSCA4Ukraine Consortium as a whole nor any individual member institutions of the MSCA4Ukraine Consortium can be held responsible for them.
The National Radio Astronomy Observatory is a facility of the National Science Foundation operated under cooperative agreement by Associated Universities, Inc. This work made use of the HPC system MEUSA, part of the Sistema Computazionale per l'Astrofisica Numerica (SCAN) of INAF-Osservatorio Astronomico di Palermo, Italy.
\end{acknowledgments}

\appendix
\section{Cooling function in the photon energy range 1.2-4.0 keV}
\label{Tycho3D:app1}

\begin{figure}
  \centering 
  \includegraphics[width=0.55\textwidth]{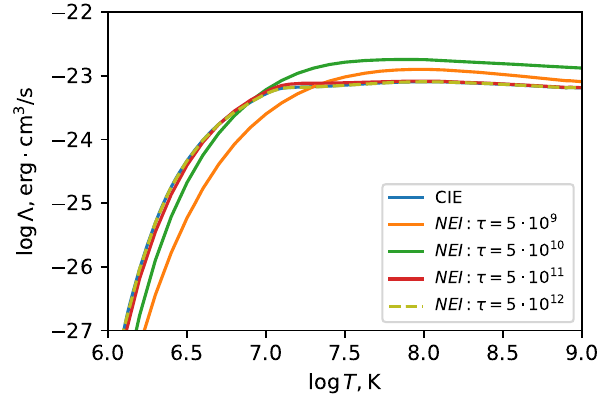}
  \hspace{0.03\columnwidth}
  \includegraphics[width=0.4\textwidth]{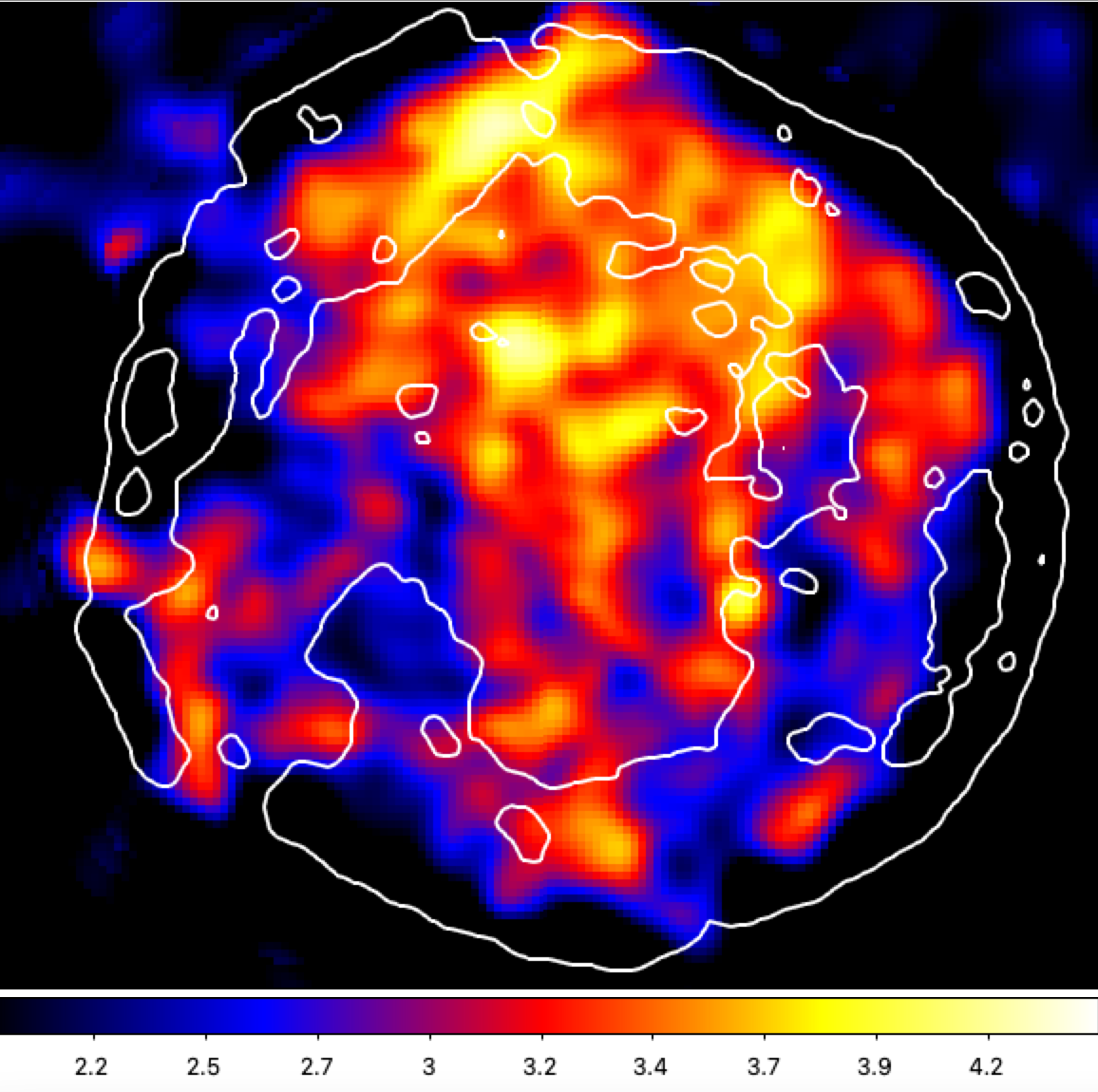}
  \caption{\textit{Left.}
    The dependence $\Lambda(T)$ in the photons with energy $1.2-4.0\un{keV}$. CIE (blue line) and NEI with values of $\tau$ shown in the legend. The line for $\tau=5\E{12}\un{cm^{-3}s}$ almost coincides with the CIE line.
    \textit{Right.} Equivalent width map for the Si emission lines in the $1.65–2.05$ keV band. The color scale is in keV. The contours correspond to the map in $4.4-6.1$ keV and are plotted at the levels $10\%$ and $30\%$ of the maximum.}
  \label{tycho3Da:fig_LambdaTconst}
\end{figure}

In order to get rid of the dependence of the cooling function $\Lambda$ on the temperature and ionization state of the plasma, we adopt the photon energy range 1.2-4 keV. 
Fig.~\ref{tycho3Da:fig_LambdaTconst} \textit{left} shows $\Lambda(T)$ for CIE as well as for NEI for a few values of the ionization parameter $\tau\equiv n\rs{o}t$. 
In order to calculate the plot, we have used the \verb+PyAtomDB+ code (\url{https://pypi.org/project/pyatomdb/}) with the abundance, which is the default in \verb+XSPEC+. 
It is evident that $\Lambda(T)\approx\mathrm{const}$ for $T>10^7\un{K}$ as in the CIE as in the NEI conditions. 
The maximum possible differences in $\Lambda$ due to local conditions are less than factor 2 (green and red lines on the figure).  
For reference, typical values of $\tau$ in Tycho SNR are $\tau\sim2\E{8}\un{cm^{-3}s}$ for thin regions of swept-up ISM around the forward shock and $\tau\sim 0.3-1\E{11}\un{cm^{-3}s}$ in the ejecta behind 
\citep{2001A&A...365L.218D,2002ApJ...581.1101H,2023ApJ...951..103E}. 

What could be an effect of spatial variation of abundance? The lines from Si and S are most prominent in the X-ray spectrum of Tycho SNR in the photon energies 1.2-4 keV. These two elements almost coincide spatially. Based on \emph{XMM-Newton} observations of the remnant, \citet{2015ApJ...805..120M} derived the equivalent width map for Si which is a good proxy for its abundance (Fig.~\ref{tycho3Da:fig_LambdaTconst} \textit{right}). This map demonstrates that variation of Si abundance is at most to the factor of two over the remnant. This is the same level of uncertainty as due to $\tau$.

%%%%%----------------------------------------%%%%%
%%%%%----------------------------------------%%%%%

%%%%%%%%%%%%%%%%%%%%%%%%%%%%%%%%%%%%%%%%%%%%%%%%%%%%%%%%
\bibliography{tychograd}{}
\bibliographystyle{aasjournal}
%%%%%%%%%%%%%%%%%%%%%%%%%%%%%%%%%%%%%%%%%%%%%%%%%%%%%%%%

\end{document}